\documentclass[fleqn,usenatbib]{mnras}

\usepackage{newtxtext,newtxmath}

\usepackage[T1]{fontenc}

\DeclareRobustCommand{\VAN}[3]{#2}
\let\VANthebibliography\thebibliography
\def\thebibliography{\DeclareRobustCommand{\VAN}[3]{##3}\VANthebibliography}

\usepackage{graphicx}	
\usepackage{amsmath}	

\usepackage{orcidlink}

\title[Joint GW PE for Pulsar Glitches]{Methods for Estimating Neutron Star Parameters using Multiple Mechanisms for Gravitational Wave Emission Associated with Pulsar Glitches}

\author[M. Ball et al.]{
Matthew Ball\,\orcidlink{0000-0001-5565-8027}\,,$^{1}$\thanks{E-mail: mball2@uoregon.edu}
Raymond Frey\,\orcidlink{0000-0003-0341-2636}\,$^{1}$\thanks{E-mail: rayfrey@uoregon.edu}
\\
$^{1}$University of Oregon, Eugene, OR 97403, USA
}

\date{Accepted XXX. Received YYY; in original form ZZZ}

\pubyear{2025}

\begin{document}
\label{firstpage}
\pagerange{\pageref{firstpage}--\pageref{lastpage}}
\maketitle

\begin{abstract}
Several mechanisms for gravitational wave (GW) emission are believed to be associated with pulsar glitches. This emission may be split between long duration continuous waves and short duration bursts. In the Advanced LIGO era, searches for GWs associated with pulsar glitches have only considered continuous wave emission. The increasing sensitivity of the detectors and the prospects for future detectors suggest that astrophysically motivated analyses involving multiple mechanisms may be possible. Here, we present a framework for combining two simple models for GW emission - long duration continuous waves and short duration bursts - to derive more constraining astrophysical implications than a single model would allow. The best limits arise from using models that predict a specific amount of GW emission; however, there are relatively few models that make such predictions. We apply these methods to the December 2016 Vela pulsar glitch and make predictions for how well future observing runs and detectors would improve results. As part of this analysis, we performed a targeted search for GW bursts associated with this glitch and find no signal.
\end{abstract}


\section{Introduction}
\label{sec:introduction}

Neutron stars are comprised of matter likely exceeding nuclear density, but the details are poorly understood. Gravitational waves (GWs) present a promising probe into this exotic matter. There are many ways a neutron star can emit GWs, but one particularly promising mechanism is a glitching pulsar \citep{link_pulsar_1992,haskell_probing_2017}.

Pulsar rotation is typically a slowly varying process, but glitches are sudden changes in the rotation rate without a clear cause \citep{fuentes_glitch_2017}. Models have been proposed \citep{haskell_models_2015}, but the precise mechanism is not understood. These models typically describe some purely internal mechanism that induces a change in angular momentum with no effect on the moment of inertia \citep{yim_transient_2020}. This would account for the lack of measurable EM emission (such as a flare) and the sudden change in the rotation rate. A rearrangement of matter capable of changing the moment of inertia could create the necessary conditions for the emission of GWs.

Pulsar glitches have been considered for gravitational wave sources since the early days of LIGO \citep{abadie_search_2011}. Modern searches are designed to separately target both long and short duration GWs using dedicated search algorithms. Long duration searches focus on GWs with durations lasting from hours to months \citep{prix_2011_search,keitel2019FirstSearch}. While all-sky, all-time GW searches are potentially capable of detecting a large signal \citep{abbott_all-sky_2021,lopez_prospects_2022,lopez_search_2024}, no targeted, short-duration search for GWs from pulsar glitches has been performed in the advanced detector era. The techniques that would be used to search for short-duration GWs from pulsar glitches would be similar to or the same as searches that target other transient neutron star phenomena such as magnetar flaring \citep[e.g.][]{macquet_search_2021,abbott_search_2024}.

Given that there is interest in both long and short duration GWs from pulsar glitches, the question may naturally arise: what information is gained by combining the search results from these emission mechanisms? Both types of emission are potentially attributable to the same type of object, so in the event of a pulsar glitch, the same source could emit both long and short duration GWs. This means that there should be an overlapping set of astrophysical parameters, and the combination of measurements from both types of search should place stronger limits on parameters of the pulsar. This paper seeks to combine measurements of both long and short duration GWs from a common pulsar glitch to place stronger constraints on the astrophysical parameters than those attainable by considering the two sets of results separately. This effect is magnified when considering models that make specific predictions of the GW energy, such as the model presented by \cite{yim_transient_2020}. Unfortunately, few models make strong predictions like this.

This paper is organized as follows. In Section \ref{subsec:search_methods}, we describe the analysis techniques that would be used in a targeted glitch search. Then, in Section \ref{subsec:GW_models}, we detail the GW waveform models we will consider for astrophysical implications. In Section \ref{sec:methods}, we describe how we use the results of a search to inform a Monte Carlo simulation to explore the parameter space permitted by the observations. In Section \ref{sec:vela2016}, we apply these methods to the 2016 Vela pulsar glitch, which occurred during the LIGO-Virgo \citep{the_ligo_scientific_collaboration_advanced_2015,acernese_advanced_2015} second observing run (O2). Finally, in Section \ref{sec:discussion_conclusions}, we consider prospects with more sensitive detectors.

The analysis we consider here consists of two distinct steps: an initial set of searches and the inference of astrophysical parameters from the search results. We give a brief overview of standard search techniques, but the main focus of this work is the interpretations step.

\section{Background}
\label{sec:background}

\subsection{Search Techniques}
\label{subsec:search_methods}

Since the two GW emission models are very different in timescale, they require different search techniques. Here, we discuss the search techniques used for each type of signal and give a more detailed description of the waveforms of interest.

The two primary waveforms take the form of slowly decaying sinusoids: a longer duration, continuous wave created by the decay of induced ellipticity or oscillations in the neutron star \citep{prix_2011_search}, and a very short duration oscillation (f-mode) created by pressure oscillations within the neutron star \citep{anderson_kokkotas-1998,keer_developing_2015,yim_gravitational_2022,wilson_gravitational_2024}. The slowly decaying waveform is expected to occur at twice the rotation rate and damp out on a longer timescale \citep{HASKELL2024102921,yim_high-priority_2024}. This model is often referred to as the ``transient mountain'' model \citep{yim_transient_2020}. The short-duration f-mode waveform is expected to occur in the range of 1000 to 3000 Hz and damp out in under a second \citep{lindblom_quadrupole_1983,mcdermott_nonradial_1988,anderson_kokkotas-1998}.

\subsubsection{Search for Continuous Waves}
\label{subsubsec:CW_search_methods}

The long-duration search for Continuous Waves (CW) integrates days of data to perform a very narrowband search around $1\times$ and $2\times$ the neutron star rotation frequency. The search pipeline we consider is the matched-filter transient $\mathcal{F}$-statistic search algorithm (implemented in \textsc{LALsuite} \citep{lalsuite}) used in the 2016 Vela glitch analysis discussed later \citep{keitel2019FirstSearch}. This algorithm is designed to report the dimensionless strain amplitude $h_0$ of the GW with a maximum likelihood \citep{jaranowski_data_1998,prix_2011_search}.

In the event of a non-detection, the pipeline places limits on the amplitude that could have been detected. Signal injection are performed for different durations to measure how well the algorithm recovers them. These injection recoveries take the form of sigmoid-like curves of the fraction of recovered waveforms at different values of $h_0$. The value of $h_0$ at $90$ or $95\%$ detection efficiency is typically reported as the upper limit. In previous searches, astrophysical implications generally take the form of estimating what fraction of some characteristic glitch energy could have been emitted as gravitational waves \citep[e.g.][]{keitel2019FirstSearch}.

\subsubsection{Search for Gravitational Wave Bursts}
\label{subsubsec:burst_search_methods}

The short-duration search for burst signals looks at multi-detector time-frequency correlations in the seconds to minutes after the glitch time, targeting GW frequencies in the range of $\sim100 \mathrm{Hz}$ to $4000 \mathrm{Hz}$. These searches are minimally modeled to account for unexpected waveforms. Two primary algorithms for this are \textsc{Coherent WaveBurst} (\textsc{cWB}) \citep{drago_coherent_2021,lopez_prospects_2022} and \textsc{X-Pipeline} \citep{sutton_x-pipeline_2010,was_performance_2012}. Both of these searches, in the case of a detection, would return time-frequency ranges of the signal with an associated integrated root-sum-square strain $h_{\mathrm{rss}}$ (Equation \ref{eq:hrss}).

\begin{equation}
    h_{\mathrm{rss}}^2 = \int_{-\infty}^{\infty}\left|\tilde{h}_{+}(f)\right|^2 + \left|\tilde{h}_{\times}(f)\right|^2 \mathrm{df}
    \label{eq:hrss}
\end{equation}

In the case of a non-detection, waveform injections are performed over a range of different signal morphologies. For the purposes of this analysis, we will only consider the ringdown waveforms (see Appendix C in \cite{abbott_search_2024}), as they are most similar to the waveforms expected for f-modes \citep{schumaker_torsional_1983,anderson_kokkotas-1998,andersson_gravitational_2003}.

Standard analyses report upper limits on the strain and compare them to the energy estimated by electromagnetic observations. For burst searches, the $h_{\mathrm{rss}}$ upper limits would be converted to an energy assuming emission via Equation \ref{eq:hrss_to_Egw} \citep{sutton_rule_2013}. This would then be compared with an estimate of the characteristic glitch energy (Equation \ref{eq:glitch_energy}) using a fiducial value for the moment of inertia (for instance $\mathcal{I}_{zz}=10^{45}\mathrm{g} \, \mathrm{cm}^2$) to estimate the relevance of the search \citep[e.g.][]{abbot_2021_allsky}.

\begin{equation}
    E_{\mathrm{GW}} = \frac{2\pi^2 c^3}{5G}d^2 f_0^2 h_{\mathrm{rss}}^2
    \label{eq:hrss_to_Egw}
\end{equation}

\subsection{GW Waveform Models}
\label{subsec:GW_models}

In order to make quantitative claims about neutron star source parameters when there is no detection, we need to make an assumption of a model for each emission mechanism. As discussed above, we consider the transient mountain model for CW emission, where the glitch induces some ellipticity that slowly damps out, and a simple damped sinusoid for the f-mode emission.

For each mechanism, we assume some multiple of a characteristic glitch energy $E_{\mathrm{Glitch}}$ is emitted in the respective channel. The characteristic glitch energy is given by Equation \ref{eq:glitch_energy} where $\mathcal{I}_{zz}$ is the moment of inertia, $\nu_{s}$ is the rotation rate of the neutron star, and $\Delta \nu_{s}$ is the change in rotation rate from the glitch. The relationship between this energy and the energy potentially emitted as GWs is highly model dependent \citep{yim_high-priority_2024}.

\begin{equation}
    E_{\mathrm{Glitch}} = 4 \pi^2 \mathcal{I}_{zz} \nu_{s} \Delta \nu_{s}
    \label{eq:glitch_energy}
\end{equation}

We treat each emission mechanism independently and model the available energy for a particular channel as some fraction $\digamma$ of the characteristic glitch energy in Equation \ref{eq:glitch_energy}. The fraction of the energy in the CW channel is $\digamma_{\!\mathrm{CW}}$ and the fraction of the energy in the f-mode channel is $\digamma_{\!\mathrm{fmode}}$. These fractions are not required to add up to $1$. In the event that the total energy of the glitch is not sufficient to explain the observations, we can let $\digamma_{\!\mathrm{CW}}$ and $\digamma_{\!\mathrm{fmode}}$ exceed unity. The total energy of the system is then:

\begin{equation}
    E_{\mathrm{GW}} = E_{\mathrm{CW}}+E_{\mathrm{fmode}} = \digamma_{\!\mathrm{CW}} E_{\mathrm{Glitch}} + \digamma_{\!\mathrm{fmode}} E_{\mathrm{Glitch}}
    \label{eq:gw_channel_energy}
\end{equation}

Whereas previous searches have used a fiducial value of $10^{45} \mathrm{g} \, \mathrm{cm}^3$ for $\mathcal{I}_{zz}$, we will use the equation of state independent relation in Equation \ref{eq:inertial_fitting_function} to relate the moment of inertia to direct mass and radius estimates (in geometric units where $a_1 = 1.317$, $a_2=-0.05043$, $a_3=0.04806$, and $a_4=-0.002692$) \citep{yagi_approximate_2017}.

\begin{equation}
    \frac{I_{zz}}{M^3} = \sum_{k=1}^{4}a_k \left(\frac{M}{R}\right)^{-k}
    \label{eq:inertial_fitting_function}
\end{equation}

\subsubsection{CW Waveform}

The transient mountain model suggests that the sudden change in rotation rate associated with the glitch induces an ellipticity $\epsilon$ which slowly damps out through the emission of GWs, initially suggested by \cite{prix_2011_search}. This amplitude is parameterized in Equation \ref{eq:mtn_model_ellip}, where $\epsilon$ is the ellipticity.

\begin{equation}
    h_{0,mtn} = \frac{4 \pi^2 G}{c^4} \mathcal{I}_{zz} \frac{(2 \nu_{s})^2}{d}\epsilon
    \label{eq:mtn_model_ellip}
\end{equation}

An amplitude can be calculated by integrating the GW luminosity as a function of time for a damped sinusoid. Equation \ref{eq:CW_amplitude} shows the max glitch energy limit and its relation to the peak amplitude of the waveform. Here, $\tau_{CW}$ is the damping time of the waveform. Previous searches for CWs near glitches from the Vela pulsar have placed upper limits on $h_{0,\mathrm{CW}}$ at approximately $1-2 \times 10^{-24}$ for signal durations of around 100 days which approaches but does not quite meet the max glitch energy limit \citep{keitel2019FirstSearch}.

\begin{equation}
    h_{0,\mathrm{CW}} = \frac{1}{2 \pi d \nu_{s}} \sqrt{\frac{5G E_{\mathrm{Glitch}}}{c^3 \tau_{CW}}}
    \label{eq:CW_amplitude}
\end{equation}

\subsubsection{f-mode Waveform}
We consider f-mode GW emission in the form of a monochromatic damped sinusoid parameterized by a frequency $\nu_{f}$ and a damping time $\tau_f$, excited at time $t=0$. This follows the parameterization of \cite{Echeverria1989} and \cite{finn1992} and is shown in Equation \ref{eq:fmode_damped_sinusoid}.

\begin{equation}
h(t) = 
    \begin{cases}
        0 & \text{for } t < 0\\
        h_{0,f} e^{-t/\tau_{f}} \sin{\left(2\pi \nu_{f}t\right)} & \text{for } t \geq 0
    \end{cases}
    \label{eq:fmode_damped_sinusoid}
\end{equation}

We can relate the f-mode frequency $\nu_{f}$ and damping time $\tau_{f}$ to the mass via the equation-of-state independent fitting functions in Equations \ref{eq:fmode_frequency} and \ref{eq:fmode_damping}, where $\overline{M} = M/1.4 M_{\odot}$ and $\overline{R} = R/10 \: \mathrm{km}$ \citep{pradhan_general_2022}.

\begin{equation}
    \nu_{f,0}(kHz)\approx0.535+1.646\left(\frac{\overline{M}}{\overline{R}^3}\right)^{1/2}
    \label{eq:fmode_frequency}
\end{equation}

\begin{equation}
\frac{1}{\tau_f(s)}\approx \frac{\overline{M}^3}{\overline{R}^4}\left[21.19-13.41\left(\frac{\overline{M}}{\overline{R}}\right)\right]
\label{eq:fmode_damping}
\end{equation}

As any targeted pulsar would have a measured rotational period, we include a slight shift of the f-mode frequency from \cite{krugerfastrotating2020}. We assume a co-rotating oscillation, as this is the ``stable'' case. For slowly rotating pulsars, this effect is almost negligible, but we include it here for completeness in Equation \ref{eq:fmode_rotation}, where $\nu_{f}$ is the frequency of the f-mode GW, $\nu_{f,0}$ is the non-rotating f-mode frequency, and $\Omega$ is the rotation frequency of the neutron star, all in units of kHz.

\begin{equation}
    \nu_f = \nu_{f,0}\left(1+0.220\left(\frac{\Omega}{\nu_{f,0}}\right)-0.0170\left(\frac{\Omega}{\nu_{f,0}}\right)^2\right)
    \label{eq:fmode_rotation}
\end{equation}

Similar to the CW case, and as calculated by \cite{owen_how_2010}, the amplitude of the f-mode is calculated by integrating the GW luminosity for a damped sinusoid (Equation \ref{eq:h0_fmode}). It is quite similar to Equation \ref{eq:CW_amplitude} as both start with the assumption of a simple damped sinusoid, but here, the f-mode frequency $\nu_f$ is not twice the rotation rate.

\begin{equation}
    h_{0,f} = \frac{1}{\pi d \nu_{f}} \left(\frac{5G}{c^3}\frac{E_{GW,f}}{\tau_{f}}\right)^{1/2}
    \label{eq:h0_fmode}
\end{equation}

To simplify things a bit, we will only be considering the transient mountain model discussed above as a potential CW emission, even though other forms of long-duration emission have been considered \citep[e.g.][]{HASKELL2024102921,yim_high-priority_2024}. We will also assume that the only short duration burst emission is the f-mode, ignoring other potential emissions.

The inclusion of $\digamma_{\!\mathrm{CW}}$ and $\digamma_{\!\mathrm{fmode}}$ explicitly gives Equations \ref{eq:mtn_amplitude_with_F} and \ref{eq:fmode_amplitude_with_F}.

\begin{equation}
    h_{0,\mathrm{CW}} = \frac{1}{2 \pi d \nu_{s}} \sqrt{\frac{5G \digamma_{\!\mathrm{CW}} E_{\mathrm{Glitch}}}{c^3 \tau_{CW}}}
    \label{eq:mtn_amplitude_with_F}
\end{equation}

\begin{equation}
    h_{0,f} = \frac{1}{\pi d \nu_{f}} \left(\frac{5G}{c^3}\frac{\digamma_{\!\mathrm{fmode}} E_{\mathrm{Glitch}}} {\tau_{f}}\right)^{1/2}
    \label{eq:fmode_amplitude_with_F}
\end{equation}

\section{Methods}
\label{sec:methods}

The long and short duration search pipelines have different result products which must be adapted to be used for parameter estimation. Here, we consider different cases for the outcomes of the different searches and how the result can be used for follow-up analysis.

\subsection{Parameter Estimation}
\label{subsec:PE}

We perform Monte Carlo simulations to generate random samples of astrophysical parameters. Since the CW and f-mode models we use should have common parameters for a common source, we have used the same prior distributions for mass, radius, glitch size, distance, the CW energy fraction $\digamma_{\!\mathrm{CW}}$, the f-mode energy fraction $\digamma_{\!\mathrm{fmode}}$, and the inclination angle $\iota$. The CW likelihood will require an additional prior for $\tau_{CW}$. 

We use Bayes' Theorem to compute a posterior distribution on the astrophysical parameters of the source pulsar. More detailed methods for how this is done in practice are given in Section \ref{subsec:joint_PE}.

\subsection{Adapting Continuous Wave Search Results}
\label{subsec:CW_adapt}

A transient CW detection will likely take the form of a posterior distribution on  $h_{0,\mathrm{CW}}$ and $\tau_{CW}$. To convert this posterior into a likelihood for follow-up analysis, we need to normalize by the prior. Bayes' theorem in the context of gravitational-wave parameter estimation takes the form

\begin{equation}
    p(\mathbf{\theta} | \mathbf{d}) = \frac{L(\mathbf{d} | \mathbf{\theta}) \pi(\mathbf{\theta})}{Z} = \frac{L(\mathbf{d} | \mathbf{\theta}) \pi(\mathbf{\theta})}{\int L(\mathbf{d} | \mathbf{\theta}) \pi(\mathbf{\theta}) d\mathbf{\theta}}
    \label{eq:bayes_theorem}
\end{equation}
where $p(\mathbf{\theta} | \mathbf{d})$ is the posterior probability density function of the source parameters $\mathbf{\theta}$ given the data $\mathbf{d}$, $\pi(\mathbf{\theta})$ is the prior probability density function, $L(\mathbf{d} | \mathbf{\theta})$ is the conditional probability density function of the data given the source parameters, and $Z$ is the evidence, which is just a renormalization factor, only dependent on the data which can be folded into the likelihood. To use an existing posterior distribution for our purposes, we need to calculate the conditional likelihood $L(\mathbf{d} | \mathbf{\theta})$. We can then treat this as any ordinary likelihood in our MCMC code. The piece we want is simply

\begin{equation}
    \frac{L_{CW}(\mathbf{d} | \mathbf{\theta})}{Z} = \frac{p_{CW}(\mathbf{\theta} | \mathbf{d})}{\pi(\mathbf{\theta})}.
    \label{eq:likelihood_from_posterior}
\end{equation}

We can perform this operation to extract the likelihood of different values of $h_{0,\mathrm{CW}}$ and $\tau_{CW}$, which we can evaluate based on computed values from the prior distribution. This likelihood function can be used with conventional MCMC analysis techniques to compute the posterior distribution for astrophysical source parameters. This likelihood function can be combined with the relevant likelihood from a burst search result to create a joint likelihood function that appropriately combines both searches.

A non-detection by a CW pipeline such as the $\mathcal{F}$-stat algorithm will be reported as upper limits on $h_{0,\mathrm{CW}}$ from injection recovery at different amplitudes and signal durations. These upper limits are computed by performing injections at different amplitudes and fitting a sigmoid curve to the recovery fractions. We can combine these detection efficiencies through simple linear interpolation to create a function that gives a detection efficiency for any set of astrophysical parameters. In the event of a non-detection, we will use this function subtracted from unity to weigh a set of sampled parameters (using the priors in Table \ref{tab:priors}) by how likely they were to not be detected by the search. This process is explained in more detail in Section \ref{sec:vela2016}.

\subsection{Adapting Burst Search Results}
\label{subsec:burst_adapt}

A burst detection (from X-Pipeline specifically) will take the form of a time-frequency box to be followed up with specific waveform consistency tests. For a narrowband signal like the damped sinusoid of an f-mode, the central frequency should correspond to the f-mode frequency, and the bandwidth should be quite small, only a few 10s of Hz at most. The duration should be comparable to the f-mode damping time.

In this case, we use a conventional noise-weighted likelihood for the f-mode model on the GW data directly. We choose an 8-second on-source window such that the trigger start time is 2 seconds into the window and an off-source window 32 times longer to compute the background spectra. Within the on-source window, we set our signal start time prior to be the trigger time from the search pipeline, and perform MCMC sampling, marginalizing over the start time using the \textsc{Bilby} framework \citep{ashton_bilby_2019} with the \textsc{Dynesty} sampling algorithm \citep{sergey_koposov_joshspeagledynesty_2024}.

A non-detection from the burst search will be reported as upper limits on the $h_{\mathrm{rss}}$ for different waveforms. Specific waveforms are injected into data at different amplitudes to determine how well the search pipeline recovers them. As discussed above, we only consider ``ringdown'' waveforms similar to the expected form of an f-mode. Each unique injection waveform is set at different frequencies and damping times to build an estimate of the parameter-dependent sensitivity. For each waveform, a detection efficiency curve is generated, representing the amplitude at which some fraction of injected waveforms were recovered. We can perform a linear interpolation of the curves at each parameter to construct a proper function of the signal parameters. With this, we can estimate the detection efficiency for any set of parameters, computing the frequency with Equation \ref{eq:fmode_frequency}, the damping time with Equation \ref{eq:fmode_damping}, and the $h_{\mathrm{rss}}$ with Equation \ref{eq:hrss}.

Since parts of the astrophysical parameter space may create frequencies or damping times outside of what was considered for the injections performed by the search pipelines, we use a nearest-neighbor approximation for parts of parameter space outside of the linear interpolation. 

\section{Example: The 2016 Vela Glitch}
\label{sec:vela2016}

Let us consider the 2016 Vela glitch as an example. At 11:38:23.188 UTC on December 12, 2016, the Vela pulsar underwent a glitch with fractional change in the rotation rate $\Delta \nu/\nu=1.431\times10^{-6}$ \citep{palfreyman_alteration_2018,basu_observed_2019,gugercinoglu_glitches_2022,basu_jodrell_2022}.\footnote{Aggregate glitch time and glitch size taken from \url{http://www.jb.man.ac.uk/pulsar/glitches.html}} This glitch occurred during the O2 and was analyzed for long-duration transient GWs \citep{keitel2019FirstSearch}. This search set upper limits at $90\%$ detection efficiency for waveforms at twice the pulsar rotation frequency with durations of up to 120 days.

Only the LIGO Hanford detector was taking data at the exact time of the glitch, so no search for short-duration GW bursts was performed; however, it seems likely that there would not have been a detection if such a search was performed. Given a fiduciary moment of inertia of $10^{45} \, \mathrm{g} \, \mathrm{cm}^2$, this glitch would signify a loss of approximately $7.07\times 10^{42} \, \mathrm{ergs}$ in rotational energy (from Equation \ref{eq:glitch_energy}). Assuming all of this energy goes into a short-duration GW burst, this would give approximately $h_{\mathrm{rss}}=7.51\times 10^{-23}$ at 1 kHz (from Equation \ref{eq:hrss_to_Egw}). This is far below the most sensitive limits set during similar short duration searches performed in O2 (see eg. \cite{abbott_all-sky_2019}).

In this section, we first perform a targeted search for short-duration GWs using only the LIGO Hanford detector with X-Pipeline and find no candidates. We then perform a similar targeted search for GWs using a window of two-detector coincidence a few hours after the glitch time in order to estimate the sensitivity of the GW network had both detectors been online.

\subsection{Targeted On-Source Burst Search}
\label{subsec:singleifo_burst}

We performed a single-detector search using the precise glitch time with the \textsc{X-Pipeline} targeted search pipeline \citep{sutton_x-pipeline_2010}. We run this on public O2 data \citep{richabbott_open_2021}. The quoted uncertainty on the glitch time was $\sigma=2.5 \, \mathrm{s}$ \citep{palfreyman_alteration_2018}, so we choose an on-source window to cover the $3\sigma$ region of $\pm 7.5 \, \mathrm{s}$. Only one candidate within the on-source window survived the data quality and consistency cuts with a p-value of 0.9, which is consistent with background. For ringdown waveforms with damping times of $0.5 \, \mathrm{s}$, we place the $90\%$ upper limits on the $h_{\mathrm{rss}}$ shown in Table \ref{tab:onsource_hrss}. Despite the poor detector antenna pattern at the time of the glitch, this result is an improvement on the limits set for the 2006 Vela pulsar glitch, which placed strain limits of $\mathcal{O}(10^{-20})$ for ringdown waveforms \citep{PhysRevD.83.042001}.

\begin{table}
    \centering
    \begin{tabular}{|c|c|}
        \hline
         Central Frequency [Hz] & $h_{\mathrm{rss,90}}\,(\mathrm{Hz}^{-1/2})$\\
         \hline \hline
         $850$ & $1.837\times10^{-21}$\\
         \hline
         $1100$ & $3.286\times10^{-21}$\\
         \hline
         $1350$ & $2.593\times10^{-21}$\\
         \hline
         $1600$ & $3.338\times10^{-21}$\\
         \hline
    \end{tabular}
    \caption{The $h_{\mathrm{rss}}$ upper limit values at $90\%$ detection efficiency for ringdown waveforms with damping times of $0.5 \, \mathrm{s}$ for the single detector on-source targeted search.}
    \label{tab:onsource_hrss}
\end{table}

\subsection{Targeted Off-Source Burst Search}
\label{subsec:new_burst}

To estimate the two-detector sensitivity near this glitch, we perform a mock search using \textsc{X-Pipeline} on a window of two-detector coincidence approximately 3.5 hours after the glitch time. This should be a reasonable estimate of what limits would have been set by a proper targeted search had both detectors been observing. 

\begin{figure}
    \centering
    \includegraphics[width=\linewidth]{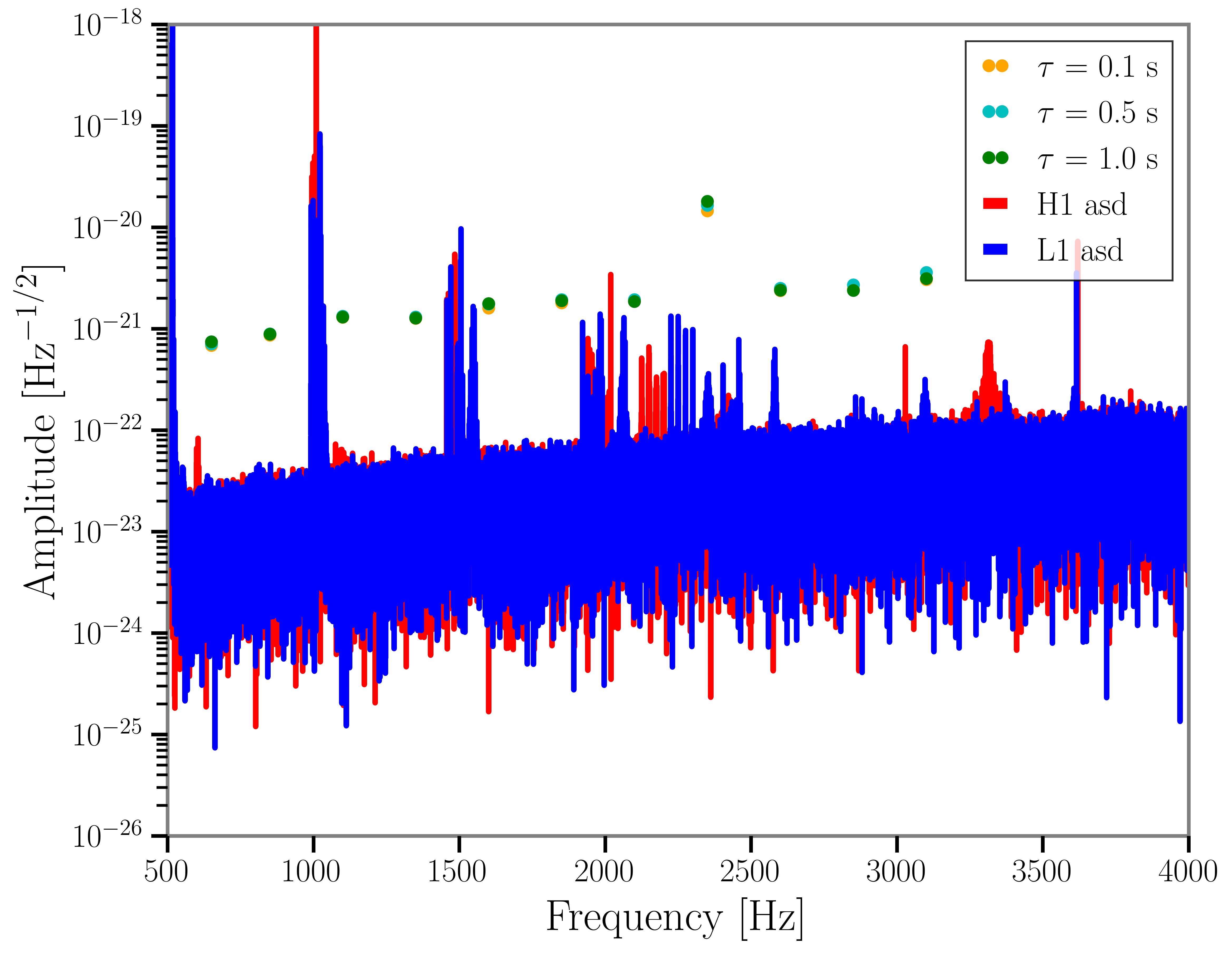}
    \caption{$90\%$ detection efficiency limits for the targeted X-Pipeline search. Ringdown injections were used to approximate sensitivity to f-mode waveforms. Upper limits are shown for damping times of $0.1\,\mathrm{s}$, $0.5\,\mathrm{s}$, and $1.0\,\mathrm{s}$. Note that for many of these points, the orange, cyan, and green dots are significantly overlapping. Background sensitivity for LIGO Hanford and LIGO Livingston are shown in red and blue, respectively.}
    \label{fig:targeted_search_limits}
\end{figure}

To estimate the sensitivity of the search, we perform injections of ringdown waveforms - damped sinusoids with a smooth ramp. We inject at frequencies ranging from 650 Hz to 3100 Hz with damping times of $0.1 \, \mathrm{s}$, $0.5 \, \mathrm{s}$, and $1.0 \, \mathrm{s}$. This should cover the majority of the potential parameter space of f-modes and capture the most important frequency and duration based effects. The $h_{\mathrm{rss}}$ value at $90\%$ detection efficiency are shown in Figure \ref{fig:targeted_search_limits}. For ringdown waveforms with damping times of $0.5 \, \mathrm{s}$, we place $90\%$ upper limits on the $h_{\mathrm{rss}}$ as shown in Table \ref{tab:offsource_hrss}. The improvement in upper limits between the two sets of results is due to using two detectors for the off-source analysis and the relative improvement in detector antenna factors that the later coincident time brought for this sky location.

\begin{table}
    \centering
    \begin{tabular}{|c|c|}
        \hline
         Central Frequency [Hz] & $h_{\mathrm{rss,90}}\,(\mathrm{Hz}^{-1/2})$\\
         \hline \hline
         $850$ & $2.547\times10^{-22}$\\
         \hline
         $1100$ & $3.640\times10^{-22}$\\
         \hline
         $1350$ & $3.310\times10^{-22}$\\
         \hline
         $1600$ & $3.489\times10^{-22}$\\
         \hline
    \end{tabular}
    \caption{The $h_{\mathrm{rss}}$ upper limit values at $90\%$ detection efficiency for ringdown waveforms with damping times of $0.5 \, \mathrm{s}$ for the off-source targeted search. This search was intended to estimate the detector network sensitivity had both interferometers been in observing at the time of the glitch. }
    \label{tab:offsource_hrss}
\end{table}

\subsection{Joint Implications}
\label{subsec:joint_PE}

We linearly interpolate the \textsc{X-Pipeline} detection efficiency curves for ringdown waveforms at multiple different frequencies and damping times to estimate the detection efficiency as a function of the frequency, the damping time, and the $h_{\mathrm{rss}}$.

We estimate the detection efficiency curves at different signal durations from the O2 long duration transient search \citep{keitel2019FirstSearch,keitel_personal}. We then interpolate these to estimate the detection efficiency of a CW as a function of the duration/damping time $\tau_{CW}$ and the amplitude $h_0$.

In order to estimate the part of parameter space permitted by the absence of a detection, we assume that a GW was emitted in both channels but not detected by either search. Under this assumption, we want the probability that the GW associated with some arbitrary set of pulsar parameters would \textit{not} be detected by a search.

We perform Monte Carlo simulations to generate random samples of pulsar parameters from the prior distributions in Table \ref{tab:priors}. For each sample, we compute the f-mode frequency $\nu_{f}$ (Equation \ref{eq:fmode_frequency}), f-mode damping time $\tau_f$ (Equation \ref{eq:fmode_damping}), and f-mode amplitude $h_{0,f}$ (Equation \ref{eq:fmode_amplitude_with_F}) to compute a waveform (Equation \ref{eq:fmode_damped_sinusoid}) and the $h_{\mathrm{rss}}$ (Equation \ref{eq:hrss}). We also compute the CW amplitude $h_{0,\mathrm{CW}}$ (Equation \ref{eq:mtn_amplitude_with_F}). The f-mode $h_{\mathrm{rss}}$, frequency and damping time are used to compute an estimated detection efficiency from the interpolated \textsc{X-Pipeline} detection efficiency curves. The CW amplitude and damping time $\tau_{CW}$ are used to compute an estimated detection efficiency from the interpolated $\mathcal{F}$-stat search detection efficiency curves. This gives a probability that this sample would be detected by each search. We subtract these probabilities from unity to get the probability that the sample would not be detected by each search.

\begin{table*}
    \centering
    \begin{tabular}{|c|c|c|}
    \hline
        Parameter & Meaning & Prior Distribution\\
        \hline \hline
        $M$ & Neutron Star Mass & $\mathrm{U}(0.9 M_{\odot},2.6 M_{\odot})$\\
        \hline
        $R$ & Neutron Star Radius & $\mathrm{U}(8 \, \mathrm{km}, \, 16\, \mathrm{km})$\\
        \hline
        $\nu_s$ & Neutron Star rotation rate & $\delta(11.19 \mathrm{Hz})$\\
        \hline
        $\Delta \nu_s / \nu_s$ & Glitch-induced spin change & $\mathcal{N}(\mu=1.431\times10^{-6},\sigma=10^{-10})$\\
        \hline
        $\iota$ & Inclination angle & $\mathcal{N}(\mu=63.6^{\circ},\sigma=0.6^{\circ})$\\
        \hline
        $d$ & Distance & $\mathcal{N}(\mu=287 \, \mathrm{pc}, \, \sigma=9.5 \, \mathrm{pc})$\\
        \hline
        $\tau_{CW}$ & CW decay time & $\mathrm{U}(0.5, 120) \, \mathrm{days}$\\
        \hline
        $\digamma_{\!\mathrm{CW}}$ & Fraction of glitch energy in CW mode & $\mathrm{logU}(10^{-3},10)$ \\
        \hline
        $\digamma_{\!\mathrm{fmode}}$ & Fraction of glitch energy in f-mode & $\mathrm{U}(0,100)$\\
        \hline
    \end{tabular}
    \caption{Prior distributions used for this analysis. The notation for these distributions is as follows: $\mathrm{U}$ refers to a uniform distribution, $\delta$ refers to a fixed value, $\mathcal{N}$ refers to a normal distribution, and $\mathrm{logU}$ refers to a log-uniform distribution. The energy scaling factors $\digamma_{\!\mathrm{fmode}}$ and $\digamma_{\!\mathrm{CW}}$ extend to values above 1 in order to demonstrate the effects of the methodology for the search limits. Additional constraints on this prior space were enforced such that the f-mode frequency is restricted to be between 800 and 3200 kHz (or the expected frequency bounds of a GW candidate), the f-mode damping time is restricted to between 0 and 10 seconds (matching the limits from the numerical simulations of \protect\cite{anderson_kokkotas-1998}), and the mass/radius values are restricted to a neutron star that obeys causality.}
    \label{tab:priors}
\end{table*}

From these probabilities, we perform rejection sampling on the initial prior samples. For each sample in the prior, we start by computing the probability of non-detection by the CW transient $\mathcal{F}$-statistic search, then probabilistically reject samples based on this likelihood. For each of the samples that remain, we compute the probability of non-detection by the \textsc{X-Pipeline} search, then probabilistically reject samples based on this second likelihood. The samples that remain represent a posterior distribution according to Bayes' Theorem.

\begin{figure}
    \centering
    \includegraphics[width=\linewidth]{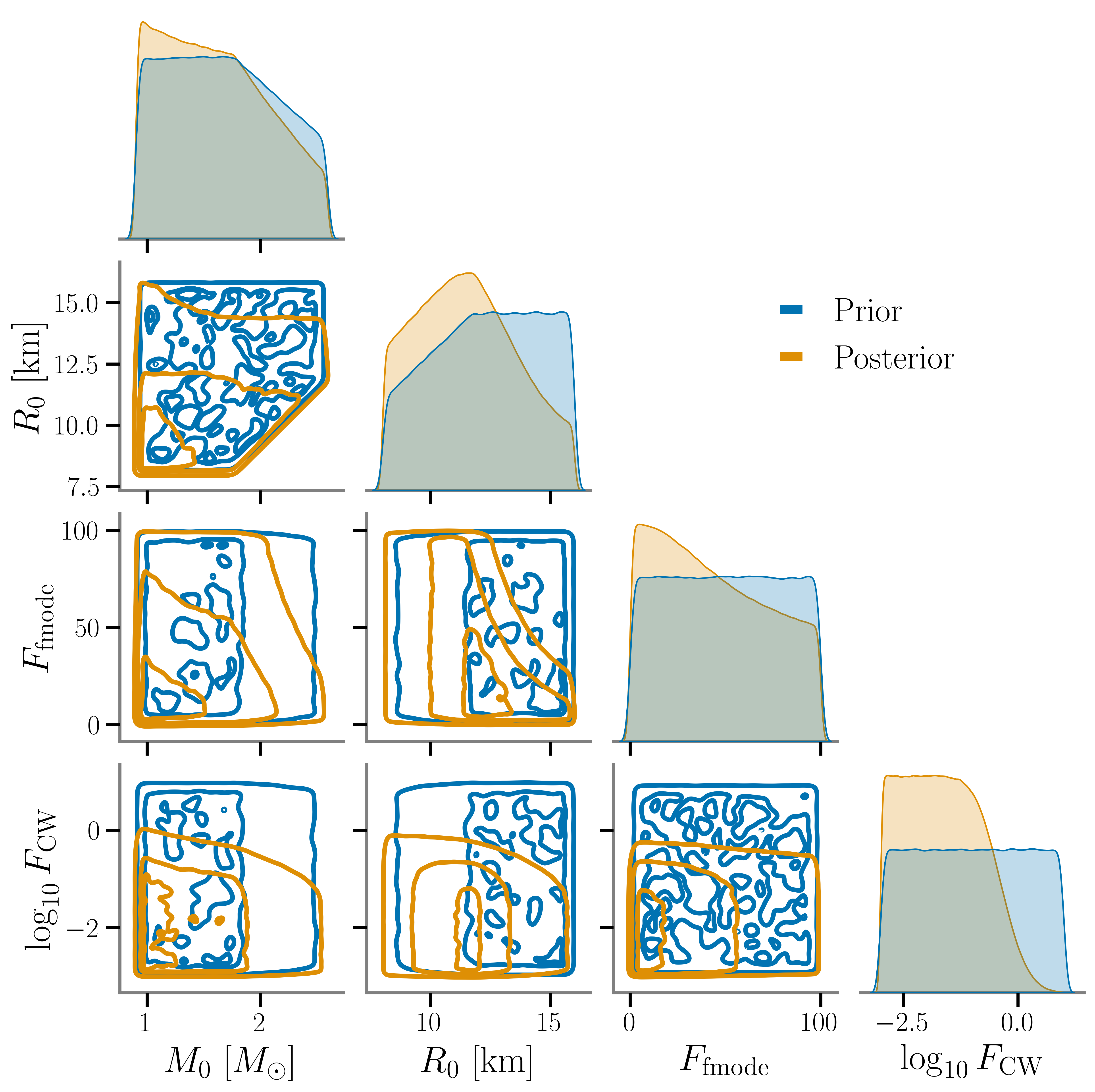}
    \caption{Mass, Radius, $\digamma_{\!\mathrm{fmode}}$, and $\digamma_{\!\mathrm{CW}}$ posterior distribution for the two-detector, off-source sensitivity. The different colors show how the prior space (blue) compares to the posterior (orange). Note that the $\digamma_{\!\mathrm{fmode}}$ and $\digamma_{\!\mathrm{CW}}$ terms are a fraction of the total glitch energy $E_\mathrm{Glitch}$, so values greater than $1$ would imply more energy than actually released in the glitch.}
    \label{fig:non_det_posterior}
\end{figure}

Figure \ref{fig:non_det_posterior} shows the posterior distribution on the mass, radius, $\digamma_{\!\mathrm{fmode}}$, and $\digamma_{\!\mathrm{CW}}$ using the transient mountain detection efficiency curves from \citep{keitel2019FirstSearch} and the ringdown detection efficiency curves from the X-pipeline search performed on a two-detector window near the glitch time (Figure \ref{fig:targeted_search_limits}).

\begin{figure}
    \centering
    \includegraphics[width=\linewidth]{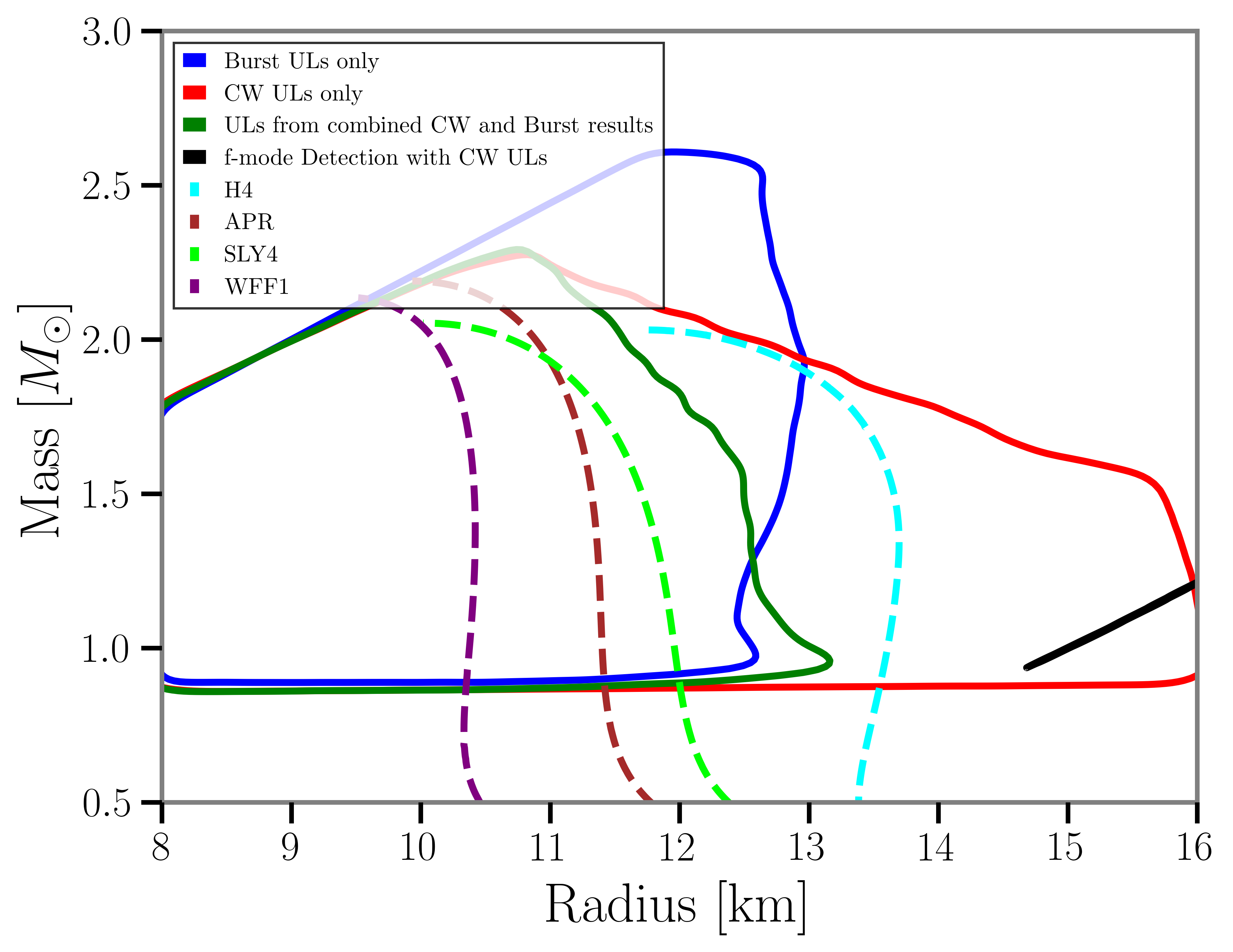}
    \caption{$90\%$ confidence region on mass and radius for different analyses. $\digamma_{\!\mathrm{fmode}}$ and $\digamma_{\!\mathrm{CW}}$ are fixed to specific values for illustrative purposes. The blue contour shows the constraints placed when only considering the upper limits from the X-Pipeline search when assuming $\digamma_{\!\mathrm{fmode}}=100$ and without considering CW results. The red contour shows the constraints placed when only considering the upper limits from the CW search when assuming $\digamma_{\!\mathrm{CW}}=1$ and without considering burst results. The green contour shows the constraints when considering both burst and CW limits when assuming $\digamma_{\!\mathrm{fmode}}=100$ and $\digamma_{\!\mathrm{CW}}=1$. Dotted lines show the mass/radius curves for different sample equations of state for reference \citep{wiringa_equation_1988,PhysRevC.58.1804,douchin_unified_2001,PhysRevD.73.024021}. Notice, that in this hypothetical example, the H4 equation of state is consistent with the confidence intervals when considering either GW model on its own. However, when combining both models, H4 is no longer consistent. The black contour shows the constraints when an f-mode is detected. Here, an f-mode was injected into detector data for a $1.0 \, M_{\odot}$ neutron star with a $15.0\,\mathrm{km}$ radius at $\digamma_{\!\mathrm{fmode}}=100$. The line-like appearance of the contour is due to the tight constraints from the f-mode frequency.}
    \label{fig:MR_contours}
\end{figure}

From this result, it is easy to see that if no detection is made when considering high values of $\digamma_{\!\mathrm{fmode}}$ and $\digamma_{\!\mathrm{CW}}$, then significant constraints could be placed on other parameters of the model. We consider how the constraints on the neutron star mass and radius might look if we assume that $\digamma_{\!\mathrm{fmode}}=100$ and $\digamma_{\!\mathrm{CW}}=1$ are fixed values. This is an extreme example meant for demonstration purposes. However, fixing these values is representative of what theoretical models might suggest (for example, the model in \cite{yim_transient_2020} suggests that $\digamma_{\!\mathrm{CW}}$ should be equal to the healing parameter $Q$ corresponding to the permanent change in the pulsar rotation rate due to the glitch). Figure \ref{fig:MR_contours} shows the $90\%$ confidence region for the mass and radius posterior distribution when making these assumptions. We consider 3 scenarios: $\digamma_{\!\mathrm{fmode}}=100$ and $\digamma_{\!\mathrm{CW}}=0$ to represent the constraints from only the f-mode upper limits, $\digamma_{\!\mathrm{fmode}}=0$ and $\digamma_{\!\mathrm{CW}}=1$ to represent the constraints from only the CW upper limits, and $\digamma_{\!\mathrm{fmode}}=100$ and $\digamma_{\!\mathrm{CW}}=1$ to represent the constraints from including both sets of limits. We also include several equations of state for reference. In this example, the H4 equation of state is permitted when only considering one model at a time but is not when considering both.

\subsection{Hypothetical f-mode Detection}
\label{subsec:burst_det}

We also consider what the results might have looked like for a hypothetical detection of an f-mode near the limits of detection. In this case, we choose a $1.0 M_{\odot}$ neutron star with a $15.0\,\mathrm{km}$ radius with $\digamma_{\!\mathrm{fmode}}=100$. Although unrealistic for a glitch of this size, this serves as a demonstration of what the follow-up analysis of a significant candidate may look like. The black contour in Figure \ref{fig:MR_contours} shows what the $90\%$ confidence region in mass-radius parameter space could look like for a detection. Note that while this appears as a line, it is actually a very narrow contour.

\section{Discussion and Conclusions}
\label{sec:discussion_conclusions}

\subsection{Future Observing Runs and Detectors}
\label{subsec:future_detectors}

In order to estimate the sensitivity of future observing runs and detectors, we wanted to project optimistic results from O2 data to future detector sensitivities. Due to the nature of the CW search, we can use the same upper limits from the original search. For the burst search, we performed an additional targeted search with X-Pipeline on O2 data but using parameters for an optimally oriented source. We chose a sky location and on-source search time to maximize the quadrature sum of the plus- and cross- polarization antenna patterns for the two LIGO detectors while having similar detector noise as the Vela targeted search performed for Section \ref{sec:vela2016}. The target window was then December 12, 2016 13:25:39 UTC with a right ascension of $0.0099718^{\circ}$ and a declination of $-39.3668^{\circ}$.

We take the $90\%$ confidence limits on $h_{\mathrm{rss}}$ (burst) and $h_0$ (CW) as our threshold for detection. We generate Monte Carlo samples from a restricted prior distribution. Here, we consider only a $1.0 M_{\odot}$ neutron star with a $15.0\,\mathrm{km}$ radius with $\digamma_{\!\mathrm{CW}}=0.1$ and $\digamma_{\!\mathrm{fmode}}=0.9$. The distance prior is uniform between 0 and $5 \, \mathrm{kpc}$ and the fractional glitch size is logarithmically uniform between $2.0\times10^{-11}$ and $1.0\times10^{-5}$ to effectively span the range of glitch sizes for exhibited by the Vela pulsar\footnote{\url{http://www.jb.man.ac.uk/pulsar/glitches.html}}.

These amplitudes are compared to the $90\%$ values, interpolated across different injections, to determine if the parameters of a particular sample meet the threshold for detection. We require samples to meet the detection threshold for both types of searches simultaneously so we can assess the potential for an ideal dual-detection.

\begin{figure}
    \centering
    \includegraphics[width=\linewidth]{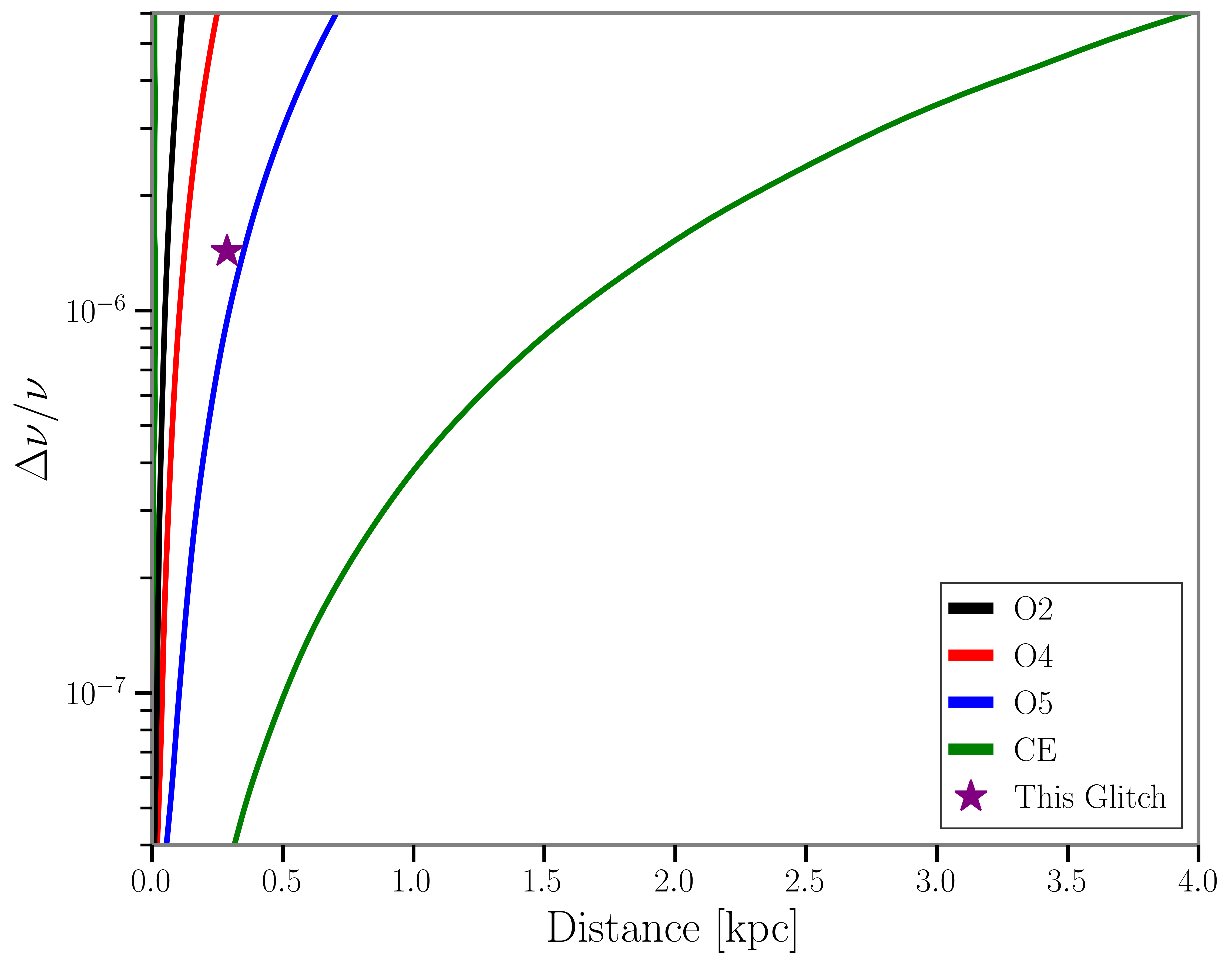}
    \caption{$90\%$ confidence interval for the distance-glitch size parameter space for a Vela-like pulsar. Glitches corresponding to values left of the lines would produce both CW transient mountain and burst f-mode GWs observable by those detectors. This assumes a $1.0 M_{\odot}$ neutron star with a $15.0\,\mathrm{km}$ radius with $\digamma_{\!\mathrm{CW}}=0.1$ and $\digamma_{\!\mathrm{fmode}}=0.9$. The purple star denotes the 2016 Vela pulsar glitch, showing that, under these assumptions, this glitch could have produced fully detectable burst and CW gravitational waves at the estimated sensitivity of O5.}
    \label{fig:future_detectors}
\end{figure}

To assess future detectors, we project these upper limits to future detectors via the ratio of the respective amplitude spectral densities at each frequency bin for each injection waveform. We compare our samples to these new thresholds to assess the dual-detectability for future detectors. Figure \ref{fig:future_detectors} shows the $90\%$ confidence interval for dual-detection in current and future observing runs and detectors. Here, we compare O2 sensitivity (black) \citep{martynov_sensitivity_2016} to O4 sensitivity (red) \citep{capote_advanced_2025}, O5 sensitivity (blue) \citep{Aplus_GWINC}, and Cosmic Explorer sensitivity (green) \citep{srivastava_science-driven_2022}. For reference, the largest glitch observed from the Vela pulsar has $\Delta \nu/\nu=3.1\times10^{-6}$ \citep{buchner_glitch_2013}, and the smallest glitch has $\Delta \nu/\nu=2\times10^{-10}$ \citep{zubieta_timing_2024}. We predict that a glitch with the properties of the 2016 pulsar glitch (purple star) could have been simultaneously detectable by both burst and CW searches at O5 sensitivity if we assume $\digamma_{\!\mathrm{CW}}=0.1$ and $\digamma_{\!\mathrm{fmode}}=0.9$.

\subsection{Benefits of Joint Analysis}
\label{subsec:benefits}

While an f-mode detection would provide precise information on the pulsar mass and radius (see Section \ref{subsec:burst_det} and Figure \ref{fig:MR_contours}), in the case of a non-detection, we can still potentially learn something new. We have demonstrated that for pulsar glitches, the non-detection of GWs associated with f-modes and transient mountains may be used to place constraints on the underlying astrophysics using the combined inference framework presented here. While this methodology can place some constraints on astrophysical parameters with a generic waveform model (Figure \ref{fig:non_det_posterior}), it is most useful when considering specific emission models that make predictions on the energetics. We emphasize that our framework for joint analysis only works under the assumption that both an f-mode and a transient mountain GWs \emph{were emitted} but were \emph{not measured} by searches. This methodology represents the first attempt - that the authors are aware of - to combine gravitational wave analyses using very different timescales. While we have found some benefit for this approach in the neutron star glitch case, we find the advantage is limited by the lack of theoretical modeling in the case of a non-detection.

Theoretical models, such as the one presented in \cite{yim_transient_2020}, make definite predictions for the energy emitted as GWs. Due to the strong dependence of the GW amplitude of our emission models on the characteristic glitch energy fractions $\digamma_{\!\mathrm{CW}}$ and $\digamma_{\!\mathrm{fmode}}$, we need predictions for these values to be able to place meaningful constraints on astrophysical parameters such as mass and radius. As an extreme example of what this may look like, we showed in Figure \ref{fig:MR_contours} the case where \emph{fictitious} models making specific predictions simultaneously resulted in much stronger constraints on the neutron star mass and radius than either of the models individually or when remaining model independent (Figure \ref{fig:non_det_posterior}). Models such as the one presented in \cite{yim_transient_2020} can be strongly tested by the framework we have presented here.

\section*{Acknowledgments}

This material is based upon work supported by NSF's LIGO Laboratory which is a major facility fully funded by the National Science Foundation. The authors acknowledge access to computational resources provided by the LIGO Laboratory supported by National Science Foundation Grants PHY-0757058 and PHY-0823459.

This research has made use of data or software obtained from the Gravitational Wave Open Science Center (gwosc.org), a service of the LIGO Scientific Collaboration, the Virgo Collaboration, and KAGRA. This material is based upon work supported by NSF's LIGO Laboratory which is a major facility fully funded by the National Science Foundation, as well as the Science and Technology Facilities Council (STFC) of the United Kingdom, the Max-Planck-Society (MPS), and the State of Niedersachsen/Germany for support of the construction of Advanced LIGO and construction and operation of the GEO600 detector. Additional support for Advanced LIGO was provided by the Australian Research Council. Virgo is funded, through the European Gravitational Observatory (EGO), by the French Centre National de Recherche Scientifique (CNRS), the Italian Istituto Nazionale di Fisica Nucleare (INFN) and the Dutch Nikhef, with contributions by institutions from Belgium, Germany, Greece, Hungary, Ireland, Japan, Monaco, Poland, Portugal, Spain. KAGRA is supported by Ministry of Education, Culture, Sports, Science and Technology (MEXT), Japan Society for the Promotion of Science (JSPS) in Japan; National Research Foundation (NRF) and Ministry of Science and ICT (MSIT) in Korea; Academia Sinica (AS) and National Science and Technology Council (NSTC) in Taiwan.

The authors would like to thank David Keitel for sharing more detailed detection efficiency curves \citep{keitel_personal} used to compute upper limits in \cite{keitel2019FirstSearch}.

The authors would like to thank Patrick Sutton for his assistance in running the \textsc{X-Pipeline} algorithm for this work.

The authors would like to thank Ian Jones and Ik Siong Heng for their valuable discussion regarding parameterization of the model used in this work. The authors would also like to thank Ian Jones and David Keitel for their insightful comments on the manuscript.

This paper has been given LIGO DCC number P2500423.
MB and RF are supported by UO NSF grant PHY-2207535.

\section*{Data Availability}

The data underlying this article are available publicly in the Gravitational Wave Open Science Center (https://gwosc.org/). The derived data used in figures will be shared on reasonable request to the corresponding author.

\bibliographystyle{mnras}
\bibliography{references}

\begin{thebibliography}{}
\makeatletter
\relax
\def\mn@urlcharsother{\let\do\@makeother \do\$\do\&\do\#\do\^\do\_\do\%\do\~}
\def\mn@doi{\begingroup\mn@urlcharsother \@ifnextchar [ {\mn@doi@}
  {\mn@doi@[]}}
\def\mn@doi@[#1]#2{\def\@tempa{#1}\ifx\@tempa\@empty \href
  {http://dx.doi.org/#2} {doi:#2}\else \href {http://dx.doi.org/#2} {#1}\fi
  \endgroup}
\def\mn@eprint#1#2{\mn@eprint@#1:#2::\@nil}
\def\mn@eprint@arXiv#1{\href {http://arxiv.org/abs/#1} {{\tt arXiv:#1}}}
\def\mn@eprint@dblp#1{\href {http://dblp.uni-trier.de/rec/bibtex/#1.xml}
  {dblp:#1}}
\def\mn@eprint@#1:#2:#3:#4\@nil{\def\@tempa {#1}\def\@tempb {#2}\def\@tempc
  {#3}\ifx \@tempc \@empty \let \@tempc \@tempb \let \@tempb \@tempa \fi \ifx
  \@tempb \@empty \def\@tempb {arXiv}\fi \@ifundefined
  {mn@eprint@\@tempb}{\@tempb:\@tempc}{\expandafter \expandafter \csname
  mn@eprint@\@tempb\endcsname \expandafter{\@tempc}}}

\bibitem[\protect\citeauthoryear{Aasi et~al.,}{Aasi
  et~al.}{2015}]{the_ligo_scientific_collaboration_advanced_2015}
Aasi J.,  et~al., 2015, \mn@doi [Classical and Quantum Gravity]
  {10.1088/0264-9381/32/7/074001}, 32, 074001

\bibitem[\protect\citeauthoryear{Abadie et~al.,}{Abadie
  et~al.}{2011a}]{PhysRevD.83.042001}
Abadie J.,  et~al., 2011a, \mn@doi [Phys. Rev. D] {10.1103/PhysRevD.83.042001},
  83, 042001

\bibitem[\protect\citeauthoryear{Abadie et~al.}{Abadie
  et~al.}{2011b}]{abadie_search_2011}
Abadie J.,  et~al., 2011b, \mn@doi [Physical Review D]
  {10.1103/PhysRevD.83.042001}, 83, 042001

\bibitem[\protect\citeauthoryear{Abbott et~al.}{Abbott
  et~al.}{2019}]{abbott_all-sky_2019}
Abbott B.,  et~al., 2019, \mn@doi [Physical Review D]
  {10.1103/PhysRevD.100.024017}, 100, 024017

\bibitem[\protect\citeauthoryear{Abbott et~al.}{Abbott
  et~al.}{2021a}]{richabbott_open_2021}
Abbott R.,  et~al., 2021a, \mn@doi [SoftwareX] {10.1016/j.softx.2021.100658},
  13, 100658

\bibitem[\protect\citeauthoryear{Abbott et~al.}{Abbott
  et~al.}{2021b}]{abbot_2021_allsky}
Abbott R.,  et~al., 2021b, \mn@doi [Phys. Rev. D]
  {10.1103/PhysRevD.104.122004}, 104, 122004

\bibitem[\protect\citeauthoryear{Abbott et~al.}{Abbott
  et~al.}{2021c}]{abbott_all-sky_2021}
Abbott R.,  et~al., 2021c, \mn@doi [Physical Review D]
  {10.1103/PhysRevD.104.122004}, 104, 122004

\bibitem[\protect\citeauthoryear{Abbott et~al.}{Abbott
  et~al.}{2024}]{abbott_search_2024}
Abbott R.,  et~al., 2024, \mn@doi [The Astrophysical Journal]
  {10.3847/1538-4357/ad27d3}, 966, 137

\bibitem[\protect\citeauthoryear{Acernese et~al.,}{Acernese
  et~al.}{2015}]{acernese_advanced_2015}
Acernese F.,  et~al., 2015, \mn@doi [Classical and Quantum Gravity]
  {10.1088/0264-9381/32/2/024001}, 32, 024001

\bibitem[\protect\citeauthoryear{Akmal, Pandharipande  \& Ravenhall}{Akmal
  et~al.}{1998}]{PhysRevC.58.1804}
Akmal A.,  Pandharipande V.~R.,   Ravenhall D.~G.,  1998, \mn@doi [Phys. Rev.
  C] {10.1103/PhysRevC.58.1804}, 58, 1804

\bibitem[\protect\citeauthoryear{Andersson}{Andersson}{2003}]{andersson_gravitational_2003}
Andersson N.,  2003, \mn@doi [Classical and Quantum Gravity]
  {10.1088/0264-9381/20/7/201}, 20, R105

\bibitem[\protect\citeauthoryear{Andersson \& Kokkotas}{Andersson \&
  Kokkotas}{1998}]{anderson_kokkotas-1998}
Andersson N.,  Kokkotas K.~D.,  1998, \mn@doi [Monthly Notices of the Royal
  Astronomical Society] {10.1046/j.1365-8711.1998.01840.x}, 299, 1059

\bibitem[\protect\citeauthoryear{Ashton et~al.,}{Ashton
  et~al.}{2019}]{ashton_bilby_2019}
Ashton G.,  et~al., 2019, \mn@doi [The Astrophysical Journal Supplement Series]
  {10.3847/1538-4365/ab06fc}, 241, 27

\bibitem[\protect\citeauthoryear{Barsotti, Fritschel, Evans  \& Gras}{Barsotti
  et~al.}{2018}]{Aplus_GWINC}
Barsotti L.,  Fritschel P.,  Evans M.,   Gras S.,  2018, {The A+ design curve},
  \url {https://dcc.ligo.org/LIGO-T1800042/public}

\bibitem[\protect\citeauthoryear{Basu, Joshi, Krishnakumar, Bhattacharya,
  Nandi, Bandhopadhay, Char  \& Manoharan}{Basu
  et~al.}{2019}]{basu_observed_2019}
Basu A.,  Joshi B.~C.,  Krishnakumar M.~A.,  Bhattacharya D.,  Nandi R.,
  Bandhopadhay D.,  Char P.,   Manoharan P.~K.,  2019, \mn@doi [Monthly Notices
  of the Royal Astronomical Society] {10.1093/mnras/stz3230}, p. stz3230

\bibitem[\protect\citeauthoryear{Basu et~al.,}{Basu
  et~al.}{2022}]{basu_jodrell_2022}
Basu A.,  et~al., 2022, \mn@doi [Monthly Notices of the Royal Astronomical
  Society] {10.1093/mnras/stab3336}, 510, 4049

\bibitem[\protect\citeauthoryear{Buchner}{Buchner}{2013}]{buchner_glitch_2013}
Buchner S.,  2013, The Astronomer's Telegram, 5406, 1

\bibitem[\protect\citeauthoryear{Capote et~al.}{Capote
  et~al.}{2025}]{capote_advanced_2025}
Capote E.,  et~al., 2025, \mn@doi [Physical Review D]
  {10.1103/PhysRevD.111.062002}, 111, 062002

\bibitem[\protect\citeauthoryear{Douchin \& Haensel}{Douchin \&
  Haensel}{2001}]{douchin_unified_2001}
Douchin F.,  Haensel P.,  2001, \mn@doi [Astronomy \& Astrophysics]
  {10.1051/0004-6361:20011402}, 380, 151

\bibitem[\protect\citeauthoryear{Drago et~al.,}{Drago
  et~al.}{2021}]{drago_coherent_2021}
Drago M.,  et~al., 2021, \mn@doi [SoftwareX] {10.1016/j.softx.2021.100678}, 14,
  100678

\bibitem[\protect\citeauthoryear{Echeverria}{Echeverria}{1989}]{Echeverria1989}
Echeverria F.,  1989, \mn@doi [Phys. Rev. D] {10.1103/PhysRevD.40.3194}, 40,
  3194

\bibitem[\protect\citeauthoryear{Finn}{Finn}{1992}]{finn1992}
Finn L.~S.,  1992, \mn@doi [Phys. Rev. D] {10.1103/PhysRevD.46.5236}, 46, 5236

\bibitem[\protect\citeauthoryear{Fuentes, Espinoza, Reisenegger, Shaw, Stappers
   \& Lyne}{Fuentes et~al.}{2017}]{fuentes_glitch_2017}
Fuentes J.~R.,  Espinoza C.~M.,  Reisenegger A.,  Shaw B.,  Stappers B.~W.,
  Lyne A.~G.,  2017, \mn@doi [Astronomy \& Astrophysics]
  {10.1051/0004-6361/201731519}, 608, A131

\bibitem[\protect\citeauthoryear{Gügercinoğlu, Ge, Yuan  \&
  Zhou}{Gügercinoğlu et~al.}{2022}]{gugercinoglu_glitches_2022}
Gügercinoğlu E.,  Ge M.~Y.,  Yuan J.~P.,   Zhou S.~Q.,  2022, \mn@doi
  [Monthly Notices of the Royal Astronomical Society] {10.1093/mnras/stac026},
  511, 425

\bibitem[\protect\citeauthoryear{Haskell}{Haskell}{2017}]{haskell_probing_2017}
Haskell B.,  2017, \mn@doi [Proceedings of the International Astronomical
  Union] {10.1017/S1743921317010663}, 13, 203

\bibitem[\protect\citeauthoryear{Haskell \& Jones}{Haskell \&
  Jones}{2024}]{HASKELL2024102921}
Haskell B.,  Jones D.,  2024, \mn@doi [Astroparticle Physics]
  {https://doi.org/10.1016/j.astropartphys.2023.102921}, 157, 102921

\bibitem[\protect\citeauthoryear{Haskell \& Melatos}{Haskell \&
  Melatos}{2015}]{haskell_models_2015}
Haskell B.,  Melatos A.,  2015, \mn@doi [International Journal of Modern
  Physics D] {10.1142/S0218271815300086}, 24, 1530008

\bibitem[\protect\citeauthoryear{Jaranowski, Królak  \& Schutz}{Jaranowski
  et~al.}{1998}]{jaranowski_data_1998}
Jaranowski P.,  Królak A.,   Schutz B.~F.,  1998, \mn@doi [Physical Review D]
  {10.1103/PhysRevD.58.063001}, 58, 063001

\bibitem[\protect\citeauthoryear{Keer \& Jones}{Keer \&
  Jones}{2015}]{keer_developing_2015}
Keer L.,  Jones D.~I.,  2015, \mn@doi [Monthly Notices of the Royal
  Astronomical Society] {10.1093/mnras/stu2123}, 446, 865

\bibitem[\protect\citeauthoryear{Keitel}{Keitel}{2024}]{keitel_personal}
Keitel D.,  2024, personal communication

\bibitem[\protect\citeauthoryear{Keitel et~al.,}{Keitel
  et~al.}{2019}]{keitel2019FirstSearch}
Keitel D.,  et~al., 2019, \mn@doi [Phys. Rev. D] {10.1103/PhysRevD.100.064058},
  100, 064058

\bibitem[\protect\citeauthoryear{Koposov et~al.,}{Koposov
  et~al.}{2024}]{sergey_koposov_joshspeagledynesty_2024}
Koposov S.,  et~al., 2024, joshspeagle/dynesty: v2.1.4,
  \mn@doi{10.5281/ZENODO.3348367}, \url
  {https://zenodo.org/doi/10.5281/zenodo.3348367}

\bibitem[\protect\citeauthoryear{Kr\"uger \& Kokkotas}{Kr\"uger \&
  Kokkotas}{2020}]{krugerfastrotating2020}
Kr\"uger C.~J.,  Kokkotas K.~D.,  2020, \mn@doi [Phys. Rev. Lett.]
  {10.1103/PhysRevLett.125.111106}, 125, 111106

\bibitem[\protect\citeauthoryear{{LIGO Scientific Collaboration}, {Virgo
  Collaboration}  \& {KAGRA Collaboration}}{{LIGO Scientific Collaboration}
  et~al.}{2018}]{lalsuite}
{LIGO Scientific Collaboration} {Virgo Collaboration}  {KAGRA Collaboration}
  2018, {LVK} {A}lgorithm {L}ibrary - {LALS}uite, Free software (GPL),
  \mn@doi{10.7935/GT1W-FZ16}

\bibitem[\protect\citeauthoryear{Lackey, Nayyar  \& Owen}{Lackey
  et~al.}{2006}]{PhysRevD.73.024021}
Lackey B.~D.,  Nayyar M.,   Owen B.~J.,  2006, \mn@doi [Phys. Rev. D]
  {10.1103/PhysRevD.73.024021}, 73, 024021

\bibitem[\protect\citeauthoryear{Lindblom \& Detweiler}{Lindblom \&
  Detweiler}{1983}]{lindblom_quadrupole_1983}
Lindblom L.,  Detweiler S.~L.,  1983, \mn@doi [The Astrophysical Journal
  Supplement Series] {10.1086/190884}, 53, 73

\bibitem[\protect\citeauthoryear{Link, Epstein  \& Van~Riper}{Link
  et~al.}{1992}]{link_pulsar_1992}
Link B.,  Epstein R.~I.,   Van~Riper K.~A.,  1992, \mn@doi [Nature]
  {10.1038/359616a0}, 359, 616

\bibitem[\protect\citeauthoryear{Lopez \& {the LIGO, Virgo, and KAGRA
  collaborations}}{Lopez \& {the LIGO, Virgo, and KAGRA
  collaborations}}{2024}]{lopez_search_2024}
Lopez D.,  {the LIGO, Virgo, and KAGRA collaborations} 2024, \mn@doi [Annalen
  der Physik] {10.1002/andp.202200142}, 536, 2200142

\bibitem[\protect\citeauthoryear{Lopez, Tiwari, Drago, Keitel, Lazzaro  \&
  Prodi}{Lopez et~al.}{2022}]{lopez_prospects_2022}
Lopez D.,  Tiwari S.,  Drago M.,  Keitel D.,  Lazzaro C.,   Prodi G.~A.,  2022,
  \mn@doi [Physical Review D] {10.1103/PhysRevD.106.103037}, 106, 103037

\bibitem[\protect\citeauthoryear{Macquet, Bizouard, Burns, Christensen,
  Coughlin, Wadiasingh  \& Younes}{Macquet et~al.}{2021}]{macquet_search_2021}
Macquet A.,  Bizouard M.~A.,  Burns E.,  Christensen N.,  Coughlin M.,
  Wadiasingh Z.,   Younes G.,  2021, \mn@doi [The Astrophysical Journal]
  {10.3847/1538-4357/ac0efd}, 918, 80

\bibitem[\protect\citeauthoryear{Martynov}{Martynov}{2016}]{martynov_sensitivity_2016}
Martynov D.~o.,  2016, \mn@doi [Physical Review D]
  {10.1103/PhysRevD.93.112004}, 93, 112004

\bibitem[\protect\citeauthoryear{McDermott, van Horn  \& Hansen}{McDermott
  et~al.}{1988}]{mcdermott_nonradial_1988}
McDermott P.~N.,  van Horn H.~M.,   Hansen C.~J.,  1988, \mn@doi [The
  Astrophysical Journal] {10.1086/166044}, 325, 725

\bibitem[\protect\citeauthoryear{Owen}{Owen}{2010}]{owen_how_2010}
Owen B.~J.,  2010, \mn@doi [Physical Review D] {10.1103/PhysRevD.82.104002},
  82, 104002

\bibitem[\protect\citeauthoryear{Palfreyman, Dickey, Hotan, Ellingsen  \&
  Van~Straten}{Palfreyman et~al.}{2018}]{palfreyman_alteration_2018}
Palfreyman J.,  Dickey J.~M.,  Hotan A.,  Ellingsen S.,   Van~Straten W.,
  2018, \mn@doi [Nature] {10.1038/s41586-018-0001-x}, 556, 219

\bibitem[\protect\citeauthoryear{Pradhan, Chatterjee, Lanoye  \&
  Jaikumar}{Pradhan et~al.}{2022}]{pradhan_general_2022}
Pradhan B.~K.,  Chatterjee D.,  Lanoye M.,   Jaikumar P.,  2022, \mn@doi [Phys.
  Rev. C] {10.1103/PhysRevC.106.015805}, 106, 015805

\bibitem[\protect\citeauthoryear{Prix, Giampanis  \& Messenger}{Prix
  et~al.}{2011}]{prix_2011_search}
Prix R.,  Giampanis S.,   Messenger C.,  2011, \mn@doi [Phys. Rev. D]
  {10.1103/PhysRevD.84.023007}, 84, 023007

\bibitem[\protect\citeauthoryear{Schumaker \& Thorne}{Schumaker \&
  Thorne}{1983}]{schumaker_torsional_1983}
Schumaker B.~L.,  Thorne K.~S.,  1983, \mn@doi [Monthly Notices of the Royal
  Astronomical Society] {10.1093/mnras/203.2.457}, 203, 457

\bibitem[\protect\citeauthoryear{Srivastava et~al.,}{Srivastava
  et~al.}{2022}]{srivastava_science-driven_2022}
Srivastava V.,  et~al., 2022, \mn@doi [The Astrophysical Journal]
  {10.3847/1538-4357/ac5f04}, 931, 22

\bibitem[\protect\citeauthoryear{Sutton}{Sutton}{2013}]{sutton_rule_2013}
Sutton P.~J.,  2013, A {Rule} of {Thumb} for the {Detectability} of
  {Gravitational}-{Wave} {Bursts}, \mn@doi{10.48550/ARXIV.1304.0210}, \url
  {https://arxiv.org/abs/1304.0210}

\bibitem[\protect\citeauthoryear{Sutton et~al.,}{Sutton
  et~al.}{2010}]{sutton_x-pipeline_2010}
Sutton P.~J.,  et~al., 2010, \mn@doi [New Journal of Physics]
  {10.1088/1367-2630/12/5/053034}, 12, 053034

\bibitem[\protect\citeauthoryear{Wilson \& Ho}{Wilson \&
  Ho}{2024}]{wilson_gravitational_2024}
Wilson O.~H.,  Ho W.~C.,  2024, \mn@doi [Physical Review D]
  {10.1103/PhysRevD.109.083006}, 109, 083006

\bibitem[\protect\citeauthoryear{Wiringa, Fiks  \& Fabrocini}{Wiringa
  et~al.}{1988}]{wiringa_equation_1988}
Wiringa R.~B.,  Fiks V.,   Fabrocini A.,  1988, \mn@doi [Physical Review C]
  {10.1103/physrevc.38.1010}, 38, 1010

\bibitem[\protect\citeauthoryear{W\k{a}s, Sutton, Jones  \& Leonor}{W\k{a}s
  et~al.}{2012}]{was_performance_2012}
W\k{a}s M.,  Sutton P.~J.,  Jones G.,   Leonor I.,  2012, \mn@doi [Phys. Rev.
  D] {10.1103/PhysRevD.86.022003}, 86, 022003

\bibitem[\protect\citeauthoryear{Yagi \& Yunes}{Yagi \&
  Yunes}{2017}]{yagi_approximate_2017}
Yagi K.,  Yunes N.,  2017, \mn@doi [Physics Reports]
  {https://doi.org/10.1016/j.physrep.2017.03.002}, 681, 1

\bibitem[\protect\citeauthoryear{Yim \& Jones}{Yim \&
  Jones}{2020}]{yim_transient_2020}
Yim G.,  Jones D.~I.,  2020, \mn@doi [Monthly Notices of the Royal Astronomical
  Society] {10.1093/mnras/staa2534}, 498, 3138

\bibitem[\protect\citeauthoryear{Yim \& Jones}{Yim \&
  Jones}{2022}]{yim_gravitational_2022}
Yim G.,  Jones D.~I.,  2022, \mn@doi [Monthly Notices of the Royal Astronomical
  Society] {10.1093/mnras/stac3405}, 518, 4322

\bibitem[\protect\citeauthoryear{Yim, Shao  \& Xu}{Yim
  et~al.}{2024}]{yim_high-priority_2024}
Yim G.,  Shao L.,   Xu R.,  2024, \mn@doi [Monthly Notices of the Royal
  Astronomical Society] {10.1093/mnras/stae1659}, 532, 3893

\bibitem[\protect\citeauthoryear{Zubieta, García, Del~Palacio, Araujo~Furlan,
  Gancio, Lousto, Combi  \& Espinoza}{Zubieta
  et~al.}{2024}]{zubieta_timing_2024}
Zubieta E.,  García F.,  Del~Palacio S.,  Araujo~Furlan S.~B.,  Gancio G.,
  Lousto C.~O.,  Combi J.~A.,   Espinoza C.~M.,  2024, \mn@doi [Astronomy \&
  Astrophysics] {10.1051/0004-6361/202450441}, 689, A191

\makeatother
\end{thebibliography}

\bsp	
\label{lastpage}
\end{document}